\newcommand{\noi}{\noindent}
\newcommand{\bc}{\begin{center}}
\newcommand{\ec}{\end{center}}

\def\ifmath#1{\relax\ifmmode #1\else $#1$\fi}
\def\3quarter{{\textstyle{3 \over 4}}}

\def\ra{\rightarrow}

\def\lf{\leaders\hbox to 1em{\hss.\hss}\hfill}

\def\21{$SU(2) \ot U(1)$}

\def\O{\hbox{$\cal O$ }}

\def\J.W.F.V{\hbox{J. W. F. Valle }}

\def\neu{\hbox{neutrino }}

\def\eq#1{{eq. (\ref{#1})}}

\def\VEV#1{\left\langle #1\right\rangle}

\def\abs#1{\left| #1\right|}

\def\lsim{\raise0.3ex\hbox{$\;<$\kern-0.75em\raise-1.1ex\hbox{$\sim\;$}}}
\def\gsim{\raise0.3ex\hbox{$\;>$\kern-0.75em\raise-1.1ex\hbox{$\sim\;$}}}

\def\bel{\begin{letter}}
\def\eel{\end{letter}}
\def\beq{\begin{equation}}
\def\eeq{\end{equation}}
\def\bef{\begin{figure}}
\def\eef{\end{figure}}
\def\bet{\begin{table}}
\def\eet{\end{table}}
\def\bea{\begin{eqnarray}}
\def\ba{\begin{array}}
\def\ea{\end{array}}
\def\bi{\begin{itemize}}
\def\ei{\end{itemize}}
\def\ben{\begin{enumerate}}
\def\een{\end{enumerate}}
\def\ra{\rightarrow}
\def\ot{\otimes}

\def\eea{\end{eqnarray}}

\def\n.c.#1#2#3{         {\it Nuovo Cim. }{\bf #1} (19#2) #3}
\def\r.n.c.#1#2#3{       {\it Riv. del Nuovo Cim. }{\bf #1} (19#2) #3}

\documentstyle[12pt]{article}
\include{matex}
\begin{document}
\begin{titlepage}
\pagestyle{empty}
\rightline{FTUV/92-39}
\rightline{IFIC/92-38}
\rightline{October 1992}
\noindent
\begin{center}
{\bf  Detection of Intermediate Mass Higgs Bosons
from Spontaneously Broken R Parity Supersymmetry}\\
\vskip 0.3cm
{\bf J. C. Rom\~ao}
\footnote{Bitnet ROMAO@PTIFM}\\
{Centro de F\'{\i}sica da Mat\'eria Condensada, INIC\\
Av. Prof. Gama Pinto, 2 - 1699 Lisboa Codex, PORTUGAL}\\
\vskip 0.3cm
{\bf J. L. Diaz-Cruz}\\
{Dpto. de Fisica, CINVESTAV-IPN\\
Ap. Postal 14-740 - 07000 Mexico DF, Mexico}\\
\vskip 0.3cm
{\bf F. de Campos}
\footnote{Bitnet CAMPOSC@EVALUN11 - Decnet 16444::CAMPOSC}\\
and \\
{\bf J. W. F. Valle}
\footnote{Bitnet VALLE@EVALUN11 - Decnet 16444::VALLE}\\
{ Instituto de F\'{\i}sica Corpuscular - C.S.I.C.\\
Dept. de F\'isica Te\`orica, Universitat de Val\`encia\\
46100 Burjassot, Val\`encia, SPAIN}
\vskip 0.3cm
{\bf ABSTRACT}\\
\end{center}
\vskip 0.1cm
\noi
The Higgs sector in spontaneously broken R Parity supersymmetry
(RPSUSY) shows interesting features that require new search
techniques. Both the mass spectrum and production rates may
differ from the standard model and minimal supersymmetric
model (MSSM) expectations. For some parameter choices,
the dominant Higgs boson decay mode can even be invisible,
leading to events with large missing transverse momentum
carried by superweakly interacting majorons. We study the
reaction $pp \ra Z + H + X$, and find that it can
lead to detectable events at LHC/SSC for a large
region of parameter space.
\noi
\end{titlepage}
\setcounter{page}{1}
\pagestyle{plain}

\section{Introduction}

Supersymmetry (SUSY) is a popular way to stabilize the
electroweak scale against quantum corrections
involving superhigh scales, thereby
addressing the so-called hierarchy problem
associated to the existence of an elementary
Higgs boson \cite{HIGGS}. As is well known,
the SUSY hypothesis requires additional elementary
Higgs bosons beyond the standard model one.

The most popular SUSY $ansatz$
is the minimal supersymmetric standard model (MSSM)
\cite{mssm} and assumes that SUSY is realized in the presence
of a discrete R parity ($R_p$) symmetry. The corresponding
Higgs sector has been extensively investigated
in the literature \cite{Hunters}.

There is great interest in investigating theories
where R parity is broken either explicitly \cite{expl}
or spontaneously \cite{MASI,ZR_RPCHI}. The latter
provides a more systematic way to include R parity
violating effects, automatically respects
low energy {\sl baryon number conservation} and
evades restrictions based on cosmological baryogenesis
arguments \cite{masiero92}
\footnote{Under special circumstances these could also
be avoided even in the explicit $R_p$ breaking scenario
\cite{dreiner}}
to the extent that the breaking of R parity sets in
spontaneously only as an electroweak scale phenomenon.
These models also require the existence of additional
Higgs bosons beyond those of the MSSM.

In this letter we analyse some of the implications for
Higgs boson physics associated with the RPSUSY model.
We concentrate on their possible impact for the upcoming LHC/SSC
hadron colliders. For definiteness we focus on the simplest
case where $R_p$ is violated in the absence of an
additional gauge symmetry beyond the minimal \21
structure \cite{MASI,pot3}. This has as distinctive
feature the existence of an associated Goldstone
boson (majoron) whose implications for Higgs
boson physics have already been noted in ref. \cite{HJJJ}.
In this model the majoron is mostly singlet,
in agreement with LEP measurements of the
invisible Z decay width \cite{LEP1}. The scale
of R-parity breaking typically lies in the
interesting range $\sim 10\:GeV-1\:TeV$,
leading to potentially large rates for the
associated $R_p$ violating effects
\cite{ROMA,RPMAJJ,MUTAU,RPMSW_RPMSWW}.

The Higgs sector of this model was studied
in a previous paper \cite{HJJJ} where the focus
was in the lightest Higgs boson and its detection
at LEP. Following that study we shall now investigate
in more detail the properties of the scalars in the model,
such as their mass spectrum and relevant couplings.
We find that these may differ substantially from the
minimal supersymmetric model (MSSM) expectations.
We concentrate on the possible detection of the
next-to-lightest scalar boson at a hadron supercollider
SSC/LHC, focusing on the likely case where its mass lies
in the intermediate mass region
\beq
\label{intmass}
m_Z < m_H  < 2m_Z
\eeq
We find that its invisible decay mode can be sizeable,
leading to events with large missing transverse momentum carried by
superweakly interacting majorons. We study the
reaction $pp \ra Z + H + X$, and find that it can
lead to detectable events at LHC/SSC for a large
region of parameter space.

\section{The Model and its Scalar Spectrum}

We consider the \21 model defined by the superpotential
terms \cite{pot3}
\bea
\label{P}
h_u u^c Q H_u + h_d d^c Q H_d + h_e e^c \ell H_d +
\hat{\mu} H_u H_d + \\\nonumber
(h_0 H_u H_d - \epsilon^2 ) \Phi +
h_{\nu} \nu^c \ell H_u + h \Phi \nu^c S +
M \nu^c S + M_\Phi \Phi \Phi + \lambda \Phi^3
\eea
The couplings $h_u,h_d,h_e,h_{\nu},h$ are described
by arbitrary matrices in generation space.
The first five terms are the usual ones that define
the $R_p$-conserving MSSM. The fifth term ensures that
electroweak breaking can take place at the tree
level, as in \cite{BFS}. The last five terms involve
\21 singlet superfields $(\Phi ,{\nu^c}_i,S_i)$
carrying lepton numbers $(0,-1,1)$ respectively
\footnote{The possible origin and phenomenological
implications of such singlets has been discussed
in refs. \cite{port_SST1_wein}.}.
In the present context their presence is important
in order to drive the spontaneous violation of R parity
and electroweak symmetries in a consistent way \cite{MASI}.
For our present purposes it will be enough to assume
that the coupling matrices ${h_{\nu}}_{ij}$ and $h_{ij}$
are nonzero only for the third generation, i.e. we set
$h_{\nu} \equiv {h_{\nu}}_{33}$ and $h \equiv h_{33}$.
With this assumption we are studying effectively
a one generation model.

To complete the specification of the model we give the
form of the full scalar potential along neutral directions
\bea
\label{V}
V_{total}  =
	\abs {h \Phi \tilde{S} + h_{\nu} \tilde{\nu} H_u }^2 +
	\abs{h_0 \Phi H_u + \hat{\mu} H_u}^2 + \\\nonumber
	\abs{h \Phi \tilde{\nu^c}}^2 +
	\abs{- h_0 \Phi H_d  - \hat{\mu} H_d +
	h_{\nu} \tilde{\nu} \tilde{\nu^c} }^2+
	\abs{- h_0 H_u H_d + h \tilde{\nu^c} \tilde{S} - \epsilon^2}^2 +
	\abs{h_{\nu} \tilde{\nu^c} H_u}^2\\\nonumber
+ \tilde{m}_0 \left[-A ( - h \Phi \tilde{\nu^c} \tilde{S}
+ h_0 \Phi H_u H_d - h_{\nu} \tilde{\nu} H_u \tilde{\nu^c} )
+ (1-A) \hat{\mu} H_u H_d \right. \\\nonumber
\left. + (2-A) \epsilon^2 \Phi + h.c. \right]
	+ \sum_{i} \tilde{m}_i^2 \abs{z_i}^2
+ \alpha ( \abs{H_u}^2 - \abs{H_d}^2 - \abs{\tilde{\nu}}^2)^2
\eea
where $\alpha=\frac{g^2 + {g'}^2}{8}$ and $z_i$ denotes any
neutral scalar field in the theory. For simplicity we have
taken $M=0$ and $\lambda=0$ from now on.

The pattern of spontaneous symmetry breaking of
both electroweak and R parity symmetries has been
studied in \cite{pot3}. There it was demonstrated
explicitly that, for suitable values of the low
energy parameters consistent with observation,
the energy is minimum when both R parity and
electroweak symmetries are spontaneously
broken. Electroweak breaking is driven by the
isodoublet VEVS $v_u = \VEV {H_u}$ and $v_d = \VEV {H_d}$,
assisted by the VEV $v_F$ of the scalar in the singlet
superfield $\Phi$. The W mass fixes the combination
$v^2 = v_u^2 + v_d^2$ while the ratio of isodoublet
VEVS defines the important parameter
\beq
\tan \beta = \frac{v_u}{v_d} \:.
\label{beta}
\eeq
On the other hand the spontaneous breaking
of R parity is driven by nonzero scalar \neu VEVS
$v_R = \VEV {\tilde{\nu^c}_{\tau}}$
and $v_S = \VEV {\tilde{S_{\tau}}}$, where
$V = \sqrt{v_R^2 + v_S^2}$ can lie anywhere in the
range $\sim 10\:GeV-1\:TeV$. A necessary ingredient for the
consistency of this model is the presence of a small
seed of R parity breaking in the $SU(2)$ doublet sector,
$v_L = \VEV {\tilde{\nu}_{L\tau}}$.

The spontaneous R parity breaking also implies
the spontaneous violation of $total$ lepton number
(conserved by \eq{P}) leading to the existence of
the majoron, given by the imaginary part of
\beq
\frac{v_L^2}{Vv^2} (v_u H_u - v_d H_d) +
              \frac{v_L}{V} \tilde{\nu_{\tau}} -
              \frac{v_R}{V} \tilde{\nu^c}_{\tau} +
              \frac{v_S}{V} \tilde{S_{\tau}}\:.
\label{maj}
\eeq
Astrophysical considerations \cite{KIM} related to
stellar cooling by majoron emission require a small
value of $v_L \sim 100 \: MeV$, which is also naturally
obtained from the minimization of the Higgs potential
\footnote{
The marked hierarchy in the values of $v_R$ and
$v_L$ follows because $v_L$ is related to a Yukawa
coupling $h_{\nu}$ and vanishes as $h_{\nu} \ra 0$.
This also leads to a \neu mass spectrum that provides
an explanation of the solar neutrino data \cite{RPMSW_RPMSWW}.}.

The scalar spectrum contains 6 physical scalars and 5 physical
pseudoscalars (one of them being the massless majoron J).
The mass eigenstates are obtained from the weak ones by the
rotations,
\beq
H_i= P_{ij} x_j \ \ \ \ \ \  A_i=Q_{ij} y_j
\label{diag}
\eeq
where $x_i,y_i$ are the real and imaginary parts
of the scalar fields. The relevant scalar boson
mass matrices were obtained from \eq{V} by
evaluating the corresponding second derivatives
at minima of the potential which break both
\21 and R-parity. The procedure has been
developed in detail in ref. \cite{pot3}.
Having found such minima we compare them with
those that break only one of these symmetries
and keep only those points for which both
electroweak and R parity breaking are
indeed energetically favored. For each of
such points we get the various Higgs boson masses
and couplings. These determine the associated
production rates at LHC/SSC and their
corresponding decay branching ratios.

For definiteness we have fixed in our
calculations some of the relevant parameters as
follows $\tan \beta=3$, $v_R=100$ GeV, $m_{top}=100$ GeV,
$m_{\phi}=1$ TeV and $v_R=v_S$. The parameter
$h_{\nu}$ was varied randomly in the region
$10^{-5} \leq h_{\nu} \leq 10^{-1}$.
The remaining parameters such as
the soft SUSY breaking gaugino mass parameter $M_2$,
the Higgsino mixing parameter $\hat \mu$,
the soft SUSY breaking scalar mass $\tilde{m}_0$
present in the MSSM are randomly varied
in the range $30 \leq M_2 \leq 250$ GeV,
$100 \leq \tilde{m}_0 \leq 1000$ GeV and
$        -250 \leq \hat \mu \leq 250$ GeV.
Finally the LEP limit on the chargino mass
$m_{\tilde{\chi^+}} \geq 45$ GeV, was implemented
and $\tilde{m}_q$ fixed at 1 TeV was
assumed in the radiative corrections.
In analogy with the MSSM, these radiative corrections
can lead to sizeable effects, such as that of raising
the mass of the lightest CP even Higgs boson, $h$,
whose values generally grow with $m_t$ and $\tan\beta$.
However, there are differences. In our model, unlike the MSSM,
even when we include radiative corrections, the decay
$h \ra A A$ is always forbidden.

For the above choice of parameters
the lightest CP even Higgs boson mass $m_h$,
is at the kinematical reach of LEP, as shown
previously \cite{HJJJ}. We have now determined
that for this significant region of parameter
space the second lightest Higgs boson mass $m_H$
lies in the so-called intermediate mass region
of \eq{intmass} while all other CP even masses
tend to be outside this region. The lightest
massive CP odd Higgs boson $A$ can lie in the
intermediate mass region of \eq{intmass} or higher.
However, for most choices of parameters,
as we will note later, it is not copiously
produced in $pp$ collisions, and may be ignored
for our present purposes.

We show in Fig. 1 the resulting regions
of $m_H$ ($m_A$) against $m_h$. Clearly, for this
choice of parameters, the lightest scalar $h$ is
in the LEP region \cite{HJJJ} and will not be
considered here. On the other hand, although
$H$ could also in this mass range (region A)
it is much more likely statistically
\footnote{In fact if we make a cut around
$80\: GeV$ we find that region A covers
only 7.5\% of the points. This is the reason
we have not considered the production and decay
of $H$ at LEP in ref. \cite{HJJJ}. Strictly
speaking $H$ could be copiously produced at
LEP and also decay invisibly. However this
is not the most likely situation.}
for it to lie in the intermediate mass region,
of \eq{intmass} (region B). Thus, it will have
to be searched with the next hadron colliders
SSC/LHC and this is the case we shall focus on.

\section{Higgs Production and Decays at LHC/SSC}

Here we summarize the Higgs production
mechanisms relevant at LHC/SSC. All of them
make use of the fact that the Higgs couples
predominantly to the heavier particles
\cite{Hunters}. We focus on electrically
neutral Higgs bosons which decay invisibly.
We therefore discard Higgs boson
production via gluon gluon fusion, the
most copious production mechanism, since
it does not lead to an easily recognizable
signal at LHC/SSC. Similarly for the
WW or ZZ fusion, that could become important
at higher energies. Finally, associated Higgs
production with $t\bar{t}$, involving
Higgs bremstrahlung from top quarks
produced in  gluon fusion, is plagued
by very large QCD backgrounds.

Thus we concentrate on associated Higgs
production with W and Z bosons,
the analogue of the Bjorken process
for hadron colliders. Although small,
these cross sections are useful to
detect intermediate mass Higgs bosons
due to the possibility of "tagging" the Higgs
by triggering on the weak intermediate vector boson.

For definiteness we focus here on the production
and decay properties of the next-to-lightest
CP even Higgs boson $H$. As mentioned above,
this is the one that most likely lies
in the intermediate mass region relevant
for LHC/SSC, given our reference choice
of parameters defined above.
The relevant lagrangian for
associated Higgs production with
the Z boson is given as \cite{HJJJ}
\beq
\label{lagz1}
{\cal L}_{Z1}=\frac{g}{2\: \cos\theta_W}M_Z\: Z_{\mu}\: Z_{\mu}\: H_2
\left[\cos\beta\: P_{21} +\sin\beta\: P_{22} +\tan\beta\: P_{23}
\right]\cos\gamma
\eeq
in terms of the next-to-lightest Higgs boson mass eigenstate
$H_2 \equiv H$.

Another production channel involves associated
Higgs production via
\beq
\label{lagz2}
{\cal L}_{Z2}=\frac{g}{2\: \cos\theta_W}Z_{\mu}\: H_2\partial^{\mu}A_j\: C_j
\eeq
where $C_j=P_{21} Q_{j1}-P_{22} Q_{j2}+P_{23} Q_{j3}$.

In \eq{lagz1} and \eq{lagz2} the $P_{ij} $ and $Q_{jk}$
coefficients come from the diagonalization of the
appropriate Higgs boson mass matrices in \eq{diag}.
We have found that the couplings $C_j$ which determine the
associated Higgs production via \eq{lagz2} are very
small (smaller than $\O (10^{-3})$ for our representative
choice of parameters. So we discard this channel
in comparison with \eq{lagz1}.

Finally, the next-to-lightest Higgs boson can be produced
in association with the W boson via \cite{HJJJ}
\beq
{\cal L}_W=g\: M_W\: W^{+}_{\mu}\: W^{-}_{\mu}\: H_2
\left[\cos\beta\: P_{21} +\sin\beta\: P_{22} +\tan\beta\: P_{23}
\right]\cos\gamma
\label{lagw}
\eeq
As we see the relevant coupling strength that
determine the Higgs production mechanisms
are $g^2_{HZZ}$ and $g^2_{HWW}$. In Fig. 2
we show the allowed region of coupling ratios
$g^2_{HZZ}/g^2_{\phi_{SM}ZZ}$
normalized with respect to the standard
model (SM) prediction, for the same previous
choice of parameters. The corresponding
ratio $g^2_{HWW}/g^2_{\phi_{SM}WW}$
is identical.

We now move to the Higgs decays. These are determined
by the relevant coupling strength parameters
$g^2_{HWW}/g^2_{\phi_{SM}WW}$,
$g^2_{HZZ}/g^2_{\phi_{SM}ZZ}$,
$g^2_{Huu}/g^2_{\phi_{SM}uu}$,
$g^2_{Hdd}/g^2_{\phi_{SM}dd}$,
$g^2_{Hll}/g^2_{\phi_{SM}ll}$,
$g^2_{HJJ}/g^2_{Htt}m^2_H$.
These in turn are determined by diagonalizing the appropriate
Higgs boson mass matrices through \eq{diag}.
By assumption, the decays into $t\bar{t}$, $WW$ and $ZZ$
are absent due to phase space. Of the remaining two-body
decays, the most likely to be important are
$H \ra b \bar{b}$ and $H \ra JJ$. While we can not
predict the relative importance of these decay branching
ratios, we have found that they can be of the same order
of magnitude for many choices of parameters.
In fact, we have determined that, for many choices
of parameters, the decay into $JJ$ can even be
the dominant $H$ decay mode. Moreover, since the
next larger branching ratio is into $hh$, and since
the dominant decay of $h$ is into $JJ$ too,
we have that the $H$ decay into invisible modes
can be the largest one. This will lead to events
characterized by a large amount of missing
transverse momentum carried by majorons.

Another useful decay for the intermediate mass range
would be, for example, $H \to Z l\bar l$, mediated by
virtual Z exchange. However, for masses below 140 GeV
this decay will not be significant. The decay into
$\gamma \gamma$ will not be taken into account
since the expected branching ratio is at most
$10^{-3}$, with little hope of a detectable signal.
We are left with the $JJ$ and $hh$ decays, as the
main decays of $H$
\footnote{It seems reasonable to neglect the
supersymmetric Higgs decay channels.}.

%

Since we are focusing on the decay of $H$ into
invisible modes, we have to search for a mechanism of
production where one has an additional particle
to tag the signal. We have at our disposal the
associated production with Z, W and $t\bar{t}$
pairs. As mentioned above, we discard its production via
gluon gluon fusion with an intermediate top loop
involving the $Ht\bar{t}$ coupling, because of the
absence of a useful tag that can make the signal
recognizable. On the other hand, the associated
Higgs production with $t\bar{t}$, involving
Higgs bremstrahlung from top quarks has too
large QCD backgrounds. Thus we shall focus
in the associated production $H+Z/W$. For
most choices of parameters in the RPSUSY model
the cross section for the production of $H+Z$
at hadron colliders can be written in terms
of the corresponding SM result via the ratio
\beq
\label{R}
R \equiv {\sigma (pp \ra H+Z+X) \over \sigma (pp \to \phi_{SM}+Z+X)} =
 g^2_{HZZ}/ g^2_{\phi_{SM} ZZ}
\eeq
The same ratio also describes associated $H+W$ production.
As seen in Fig. 2 the ratio of the couplings
above can reach values of \O(1), making the
corresponding cross-section comparable to the
known SM value. For the signature of the events
we shall include the decay of Z into $l^+l^-$
($l=e, \mu$) and the invisible decay of $H$.
This leads to events involving a pair of leptons
plus large missing transverse momentum. The dominant
background will come from the production of a pair ZZ,
with one of them decaying into charged leptons and the other
into neutrinos, which will fake the invisible decay of $H$.

This has recently been considered by Kane and collaborators
\cite{kane}, who have studied the detection of a Higgs
boson that decays into invisible modes. In contrast
to the spirit of the present paper, these authors
have considered the question in a general
unspecified theoretical context. Their results are
applicable to us and can be
used immediately in order to assess the potential
detectability of the above signature in our model.
{}From ref. \cite{kane} it is concluded that this kind
of signal can be detected for a Higgs mass as large as
160 GeV, provided that two assumptions are obeyed, namely that
\ben
\item
the coupling $HZZ$ is of the same strength as in the SM  and
\item
the decay into invisible modes has 100 $\%$ branching
\een
As we have shown in the present model both the
$H$ decay branching ratio and the $HZZ$ coupling that
determine its production rate can be \O(1).
Other situations for which these quantities are
smaller can be considered by appropriately
rescaling from the results corresponding
to the above reference point. Correspondingly the discovery
potential for this signal will be weaker.

If one adopts as a discovery criteria four
standard deviations of signal above background,
then one can evaluate the minimum number of
events that should have the signal, and
correspondingly determine the minimum value of
$f= g^2_{HZZ}/g^2_{\phi_{SM}ZZ} BR(H \to invisible)$.
Using the
results of Kane et al. we can determine the minimum
value of $f$ required (for several values of $m_H$) to
have a $4\sigma$ signal. This is shown in table 1.
By comparing this table with Fig. 2 one sees that,
Higgs masses up to 140 GeV or so can be reached
at LHC/SSC and that these are meaningful
theoretically in the present model.

To conclude this section we note that the associated
production with W would also lead to interesting events
where the W gives an $e$ or $\mu$ and the Higgs again
decays invisibly. However this has only one lepton
for reconstruction which can make the detection
more difficult. As noted in \cite{kane} it could
however be used to reconfirm a possible signal in
the Z channel.

\section{ Conclusions}

We have studied the Higgs sector of the
spontaneously broken R-parity model of
refs. \cite{MASI} and \cite{pot3}. We
showed that the mass spectrum of this model
favors the existence of a Higgs boson $H$ in
the intermediate mass range. Its decay into
invisible modes, either directly or via
$H \ra hh$, where $h$ is the lightest CP
even Higgs boson is substantial, and can
even be the dominant one.
Our main conclusion is that in
this RPSUSY model $H$ could be discovered at
SSC/LHC using the associated production with Z
for values of the parameters that are both
theoretically consistent and experimentally
allowed. The associated production with W would also
lead to interesting events where the W gives an $e$
or $\mu$ and the Higgs again decays invisibly. This
could be used to provide a reconfirmation of a possible
signal in the Z channel.\\
\noi
This work was supported by CICYT (Spain) and by a CNPq
a fellowship (Brazil).

\begin{table}
\begin{center}
%
%
\begin{tabular}{|l|c|c|} \hline
$m_{H}$(GeV) & $f^{ZH}_{min}$-SSC(LHC) & $f^{WH}_{min}$-SSC(LHC) \\ \cline{1-3}
40 & $4\times10^{-2}(5.2\times10^{-2})$ & $1\times10^{-2}(1.4\times10^{-2})$ \\
50 & $6\times10^{-2}(7.5\times10^{-2})$ & $1.6\times10^{-2}(2\times10^{-2})$ \\
60 & $8\times10^{-2}(1.1\times10^{-1})$ & $2.3\times10^{-2}(3\times10^{-2})$ \\
70 & $1.2\times10^{-1}(1.5\times10^{-1})$ &
$3.3\times10^{-2}(4.3\times10^{-2})$ \\
80 & $1.4\times10^{-1}(2\times10^{-1})$ & $4.4\times10^{-2}(5.9\times10^{-2})$
\\
90 & $1.9\times10^{-1}(2.7\times10^{-1})$ &
$5.8\times10^{-2}(7.8\times10^{-2})$ \\
100 & $2.4\times10^{-1}(3.4\times10^{-1})$ & $7.5\times10^{-2}(1\times10^{-1})$
\\
110 & $3\times10^{-1}(4.3\times10^{-1})$ & $9.6\times10^{-2}(1.4\times10^{-1})$
\\
120 & $3.9\times10^{-1}(5.5\times10^{-1})$ &
$1.3\times10^{-1}(1.9\times10^{-1})$ \\
130 & $4.9\times10^{-1}(6.9\times10^{-1})$ &
$1.9\times10^{-1}(2.8\times10^{-1})$ \\
140 & $6\times10^{-1}(8.9\times10^{-1})$ & $3.2\times10^{-1}(4.7\times10^{-1})$
\\ \hline
\end{tabular}
\end{center}
\caption{Minimun value of
$f= g^2_{HZZ}/g^2_{\phi_{SM}ZZ} BR(H \to invisible)$
which would be required for a four standard deviation
signal in the channels $ZZH$ and $WWH$ at SSC(LHC) for
Higgs masses up to $140\: GeV$. The signal could be
detectable for masses below this value. }
\end{table}
\newpage
{\bf \large Figure Captions}\\

{\bf Fig. 1}
Regions A and B in Figure 1a show allowed values for
the mass of the next-to-lightest scalar Higgs boson
$H$ versus that of the lightest scalar $h$ in the
RPSUSY model. Region A in Figure 1b shows the
corresponding allowed values for the pseudoscalar
mass $m_A$ versus $m_h$. We have fixed $\tan\beta=3$,
$m_{top}=100\; GeV$ and $v_{R}=100\; GeV$.
All of these regions are allowed by experiment and by
the minimization of the scalar potential of the theory.
However, there are many more choices of parameters that
lead to points in region B than in A of Figure 1a.\\

{\bf Fig. 2}
Region A in Figure 2 shows the allowed values for the squared
$ZZH$ coupling, normalized with respect to that of the SM, for
the same parameter choice as in Figure 1. As can be seen from
the figure, in the RPSUSY model this coupling can attain,
as its maximum allowed value, the canonical SM prediction.\\

\newpage

\begin{thebibliography}{10}

\bibitem{HIGGS}
 P.~W. Higgs,
\newblock {\em Phys.\ Lett.} {\bf 12}, 132 (1964).

\bibitem{mssm}
 H. Haber and  G. Kane,
\newblock {\em Phys.\ Rev.} {\bf 117}, 75 (1985).

\bibitem{Hunters}
 S. Dawson etal, {\em The Higgs Hunters Guide},
(Addison Wesley, 1990)

\bibitem{expl}
 L. Hall and  M. Suzuki,
\newblock {\em Nucl.\ Phys.} {\bf B231}, 419 (1984).

\bibitem{MASI}
 A. Masiero and  J. W.~F. Valle,
\newblock {\em Phys.\ Lett.} {\bf B251}, 273 (1990).

\bibitem{ZR_RPCHI}
 J. W.~F. Valle,
\newblock {\em Phys.\ Lett.} {\bf B196}, 157 (1987);
 M.~C. Gonzalez-Garcia and  J. W.~F. Valle,
\newblock {\em Nucl.\ Phys.} {\bf B355}, 330 (1991).

\bibitem{masiero92}
 G.F.Giudice, A. Masiero, M. Pietroni and A. Riotto,
CERN-TH6656/92(1992).

\bibitem{dreiner}
 H.~Dreiner, G. G. Ross,
\newblock Oxford preprint OUTP-92-08P (1992).

\bibitem{pot3}
 J.~C. Rom\~ao, C.~A. Santos, and  J. W.~F. Valle,
\newblock {\em Phys.\ Lett.} {\bf B288}, 311 (1992).

\bibitem{LEP1}
 J. Steinberger,
\newblock in {\em Electroweak Physics Beyond the Standard Model},  ed.\  J.
  W.~F. Valle and  J. Velasco (World Scientific, Singapore, 1992), p.3.

\bibitem{ROMA}
 P. Nogueira,  J.~C. Rom\~ao, and  J. W.~F. Valle,
\newblock {\em Phys.\ Lett.} {\bf B251}, 142 (1990).

\bibitem{RPMAJJ}
 M.~C. Gonzalez-Garcia,  J.~C. Rom\~ao, and  J. W.~F. Valle,
\newblock {\em Nucl.\ Phys.} {\bf B} (1992)
\newblock , in press; Valencia preprint FTUV/91-42.

\bibitem{MUTAU}
 J.~C. Rom\~ao, N. Rius, and  J. W.~F. Valle,
\newblock {\em Nucl.\ Phys.} {\bf B363}, 369 (1991).

\bibitem{RPMSW_RPMSWW}
 J.~C. Rom\~ao, and  J. W.~F. Valle,
\newblock {\em Phys.\ Lett.} {\bf B272}, 436 (1991);
\newblock {\em Nucl.\ Phys.} {\bf B381}, 87 (1992).

\bibitem{HJJJ}
 J.C. Rom\~ao, F. de Campos and J.W.F.Valle,
\newblock {\em Phys.\ Lett.} {\bf B292}, 329 (1992).

\bibitem{BFS}
 R. Barbieri,  S. Ferrara, and  C. Savoy,
\newblock {\em Phys.\ Lett.} {\bf B119}, 343 (1982).

\bibitem{port_SST1_wein}
 J. W.~F. Valle,
\newblock {\em Nucl.\ Phys. \ Proc. \ Suppl.} {\bf 11}, 118 (1989).
 R. Mohapatra and  J. W.~F. Valle,
\newblock {\em Phys.\ Rev.} {\bf D34}, 1642 (1986).

\bibitem{KIM}
 J.~E. Kim,
\newblock {\em Phys.\ Rep.} {\bf 150}, 1 (1987).

\bibitem{Haber91}
 J. Ellis et~al.,
\newblock {\em Phys.\ Lett.} {\bf B262}, 477 (1991);
 H. Haber et~al.,
\newblock {\em Phys.\ Rev.\ Lett.} {\bf 66}, 1815 (1991);
 R. Barbieri et~al.,
\newblock {\em Phys.\ Lett.} {\bf B258}, 395 (1991).

\bibitem{kane}
S. Frederiksen, N. Johnson, G. Kane, and J. Reid,
\newblock SSCL-preprint-577 (1992).

\end{thebibliography}

\end{document}